\documentclass{article}
\usepackage{graphicx}
\usepackage{bm}
\usepackage{amssymb}
\usepackage{cite}
\def\fl{$f(\textrm{Lovelock})$}
\def\bea{\begin{equation}}
\def\eea{\end{equation}}

\def\beq{\begin{eqnarray}}
\def\eeq{\end{eqnarray}}
\def\LL{Lanczos-Lovelock}
\author{ Sudipta Sarkar\footnote{sudiptas@iitgn.ac.in}
\\ \textit{Indian Institute of Technology, Gandhinagar}\\
 \textit{Visat-Gandhinagar Highway, Chandkheda}\\\textit{ Ahmedabad, Gujarat 382424, India}\\ \\Aron C. Wall\footnote{aroncwall@gmail.com}
\\ \textit{Department of Physics} \\ \textit{University of California, Santa Barbara}
\\ \textit{Santa Barbara, CA 93106, USA}}
\title{Generalized second law at linear order for actions that are functions of Lovelock densities}

\date{\today}

\begin{document}

\maketitle
\begin{abstract}
In this article we consider the second law of black holes (and other causal horizons) in theories where the gravitational action is an arbitrary function of the Lovelock densities.  
We show that there exists an entropy which increases locally, for linearized perturbations to regular Killing horizons.  
In addition to a classical increase theorem, we also prove a generalized second law for semiclassical, minimally-coupled matter fields.
\end{abstract}
\section{Introduction}

General relativity is a nonrenormalizable quantum theory.  But at energies much lower than the Planck scale, it can be approximated with a low-energy effective field theory, working perturbatively in powers of Newton's constant $G$.  This effective field theory consists of general relativity plus higher-curvature counterterms arising from loop corrections.


In principle, all possible terms consistent with diffeomorphism invariance may appear at the level of the effective Lagrangian. Still, from a purely phenomenological point view, a subset of terms which leads to well-behaved classical physics are more desirable.  One possible criterion is to consider those higher curvature terms which retain the essential features of general relativity.

A guiding principle comes from the study of black hole physics which has provided strong hints of a deep relationship between gravitation, thermodynamics and quantum theory. The thermodynamic behavior of black holes \cite{Bardeen:1973gs} and other causal horizons \cite{Gibbons:1977mu,Jacobson:2003wv} in general relativity has suggested many insights into the nature of quantum phenomena in strong gravitational fields.  The core result is the identification of one quarter of the horizon area with the entropy associated with the horizon.  This area-entropy relationship in general relativity is mainly motivated from the second law of black hole mechanics.  But do the laws of black hole mechanics (in particular, the second law) generalize to higher-curvature gravity?

In this article, we will discuss the second law for actions which are functions of Lovelock densities. We show that at least for linearized perturbations, there exists an entropy which obeys a local increase law, for classical field sources obeying the null energy condition.  Then, for semiclassical perturbations by quantum fields, we will show how to obtain the generalized second law.

The paper is organized as follows: In section \ref{SL_GR}, we review the classical and semiclassical second law for general relativity.  In section \ref{HC}, we will discuss how to define the entropy for higher curvature gravity, and compare our new results to what was already known.  Next, in sections \ref{csl} and \ref{gsl}, we present the essential conditions required to prove a quasistationary classical second law as well as semiclassical generalized second law, for any theory of higher-curvature gravity.  Section \ref{fL} contains the proof of the key condition for \LL\ and f(Lovelock) theories. Finally, we conclude in section \ref{dis} by discussing the limitations of the result, and prospects for future research.

\section{The Second Law in General Relativity}\label{SL_GR}

Let us start by reviewing the situation in general relativity.  It is well known that causal horizons (e.g. black holes, de Sitter, Rindler) have thermodynamic properties \cite{Bardeen:1973gs, Gibbons:1977mu,Jacobson:2003wv}.  In particular, there is a second law which states that an ``entropy'' is increasing with time; in general relativity, the entropy of a horizon $H$ is proportional to its area:
\begin{equation}\label{Bek}
S_H = \frac{A}{4\hbar G}.
\end{equation}

The Classical Second Law (CSL) of general relativity states that so long as (i) the matter fields obey the null energy condition $T_{ab} k^a k^b \ge 0$ ($k^a$ being null), and (ii) there are no naked singularities on $H$, then this area is increasing as a function of time  \cite{Hawking:1971vc}.  The proof uses the focussing properties of the Raychaudhuri equation.  If we choose $\lambda$ to be an affine parameter, then the expansion $\theta = \frac{1}{A}\frac{dA}{d\lambda}$ obeys the Raychaudhuri Eq:
\begin{equation}\label{Ray}
\frac{d\theta}{d\lambda} = -\frac{\theta^2}{D-2} - \sigma_{ab} \sigma^{ab} - R_{ab} k^a k^b,
\end{equation}
where $\sigma_{ab}$ is the shear and the whole right-hand-side is negative.  

The increase theorem applies locally: at every point on $H$ the expansion $\theta \ge 0$ (due to Eq. (\ref{Ray}) and the future boundary condition).  It also applies to highly nonlinear processes such as the merger of two black holes.  In nonlinear processes, new generators can enter a causal horizon (but can never exit) and this also only increases the area.

In semiclassical general relativity minimally coupled to matter fields, there is a quantum analogue known as the Generalized Second Law (GSL).  The \emph{generalized entropy} is $S_\mathrm{gen} = S_H + S_\mathrm{out}$, the second term being the entropy of any matter fields outside of $H$ (see \cite{Wall09} for a review).\footnote{One must renormalize in order to absorb ultraviolet divergences of $S_\mathrm{out}$ into counterterms in $S_H$ \cite{Susskind94, Jacobson94, Demers95, Kabat95, deAlwis95, Larsen96, Frolov97, Frolov98}.  The leading order divergence corresponds to renormalizing $G$, while the subleading terms seem to correspond to adding higher-curvature terms to the action.  This seems to work for scalars and spinors, although there are discrepancies for gauge fields \cite{Kabat95, Fursaev97}.  These counterterms are higher order in $\hbar$ and can therefore be neglected at leading order in an $\hbar$ expansion.  However, they motivate the extension of horizon thermodynamics to general higher-
curvature gravity theories \cite{Jacobson:1993vj}.}

In the semiclassical approximation used by Ref. \cite{Wall:2011hj}, one expands out the metric in powers of the Planck length; if $G$ is fixed, one can think of this as an $\hbar$ expansion:
\begin{equation}
g_{ab} = g_{ab}^0 + g_{ab}^{1/2} + g_{ab}^1 + \mathcal{O}(\hbar^{3/2})
\end{equation}
Here the zeroth order term corresponds to the classical background metric, the half order term comes from gravitons quantized on the background metric, and the first order term comes from the gravitational fields produced by matter and gravitons.  Note that if the classical background has an increasing area, the CSL implies the GSL.  Furthermore, Eq. (\ref{Ray}) implies that there is no half-order component to $\theta$, although there may be to $\sigma_{ab}$. 

Therefore, the interesting questions have to do with the $\hbar^1$ part of the metric.  If one chooses to think of $\sigma_{ab} \sigma^{ab}$ as the gravitational contribution to the null stress-energy tensor $T_{ab} k^a k^b$, then the GSL arises from a linear equation:
\begin{equation}\label{linRay}
\frac{d\theta}{d\lambda} = -8\pi G\,T_{ab}
\end{equation}
This equation gives the \emph{linearized} response of a classically stationary black hole metric to a quantum perturbation.  Assuming (i) the existence of a suitable renormalization scheme, and (ii) that the matter fields satisfy certain axioms (which can be explicitly verified for free fields of various spins, and certain superrenormalizable interactions), it follows from Eq. (\ref{linRay}) that $\delta S_\mathrm{gen} \ge 0$ in a similarly local way \cite{Wall:2011hj}.  More details of this argument will be provided in section \ref{gsl}.

\section{Higher Curvature Gravity}\label{HC}

\subsection{The Entropy and its Ambiguities}

We want to know how this story generalizes to higher-curvature theories of gravity.  The challenge here is that $S_H$ is no longer given by (\ref{Bek}); there are corrections.  The standard formula used to calculate the ``Wald entropy'' of a stationary Killing horizon is \cite{Wald:1993nt, Iyer:1994ys, Jacobson:1993vj}
\begin{equation}\label{wald}
S_W = - 2\pi \int_s d^{D-2}y\,\frac{\partial L}{\partial R_{abcd}} \epsilon_{ab} \epsilon_{cd},
\end{equation}
where $L$ is the Lagrangian, $s$ is a cross-section of the horizon, $\epsilon_{ab}$ is the binormal to the horizon, and $d^{D-2}y$ is the metrized area measure over the transverse coordinates, labelling horizon generators.\footnote{If the entropy change is evaluated on a noncompact horizon, we will assume that the initial and final slices coincide except in a compact region.  This helps ensure that the entropy change is finite, and also allows total derivatives to be dropped.}  It is also possible to use this formula for first order variations away from a Killing horizon, but only when $s$ coincides with a regular bifurcation surface $B$.  This entropy has been proven to obey a stationary comparison version of the First Law \cite{Wald:1993nt, Iyer:1994ys}.

However, it is important to emphasize that for nonstationary black holes, this Wald entropy may not be valid.  It was only ever derived up to certain ambiguities in the Noether charge \cite{Iyer:1994ys, Jacobson:1993vj}, which take the general form
\begin{equation}\label{ambi}
\Delta S_H = \int_s d^{D-2}y\, V \cdot W
\end{equation}
where $V$ and $W$ are objects which transform nontrivially under boosting the normal directions, although their product is boost invariant.  These ambiguities vanish for Killing horizons, and at first order when $s = B$, but they do not vanish when $s \ne B$, even at first order.  Hence the ambiguities matter when attempting to prove a local form of the second law.

For example, for a theory of metric gravity which is quadratic in $R_{abcd}$, dimensional analysis says that $S_H$ should be given by the integral of quantities with weight 2.  There are therefore two possible ambiguity terms corresponding to the two ways to contract the extrinsic curvature $K_{ab}^i$ with itself (here $a,b$ are indices in the D-2 transverse directions, and $i$ is an index in the 2 normal directions).  Hence in this case the entropy would take the form:
\begin{equation}
S_H = - 2\pi \int_s d^{D-2}y\, \left[ \frac{\partial L}{\partial R_{abcd}} \epsilon_{ab} \epsilon_{cd}
 + c_1 K_{ab}^i K^{ab}_i + c_2 K_a^{ai} K_{bi}^b \right].
\end{equation}
where $c_1$ and $c_2$ are coefficients which cannot be determined by the First Law, but must be correctly chosen if there is to be a local second law.  Although the null extrinsic curvature along the horizon (corresponding to $\theta$ and $\sigma_{ab}$) vanishes on the stationary background, the null extrinsic curvature of null surfaces falling into the horizon does not vanish on a typical slice $s \ne B$.  As a result, the $(K_{ab}^{i})^2$ terms do not vanish at first order.  

Theories of gravity with additional powers of $R_{abcd}$ (or its derivatives) might have ambiguities involving more derivatives or more powers of the metric.  Note that any ambiguity involving four or more powers of the extrinsic curvature $K_{ab}^i$ cannot be fixed by considerations involving the linearized second law, since it would be quadratic in $\theta$ and $\sigma_{ab}$.

Next, we would like to discuss the choice of horizon entropy $S_H$ for various higher curvature theories.

\subsection{f(R) gravity}

In the case of f(R) gravity, or a nonminimally coupled scalar field, one can perform a field redefinition to transform the theory into pure general relativity minimally coupled to matter \cite{Whitt:1984pd, Magnano:1987zz}.  By means of this argument, or a direct proof \cite{Jacobson:1993vj}, one can show that the CSL also applies to these theories.  In this case the Wald formula (\ref{wald}) is correct, and the entropy is a function of the $D$-dimensional Ricci scalar at that point:
\begin{equation}
S_H = S_W = \frac{1}{4 \hbar G} \int_s d^{D-2}y\,f'( ^{(D-2)}\!R )
\end{equation}

\subsection{\LL\ gravity}

The next interesting case to check is \LL\ gravity, the most general metric theory of gravitation whose equations of motion are second order in derivatives.  Its action is given by the sum of dimensionally extended Euler densities \cite{Lanczos-Lovelock:1971yv},
\beq 
I =  \int d^Dx \sqrt{g} \sum \limits_{m=0}^{[D-1)/2]} \alpha_{(m)} {\cal L}_{(m)}^D,
\eeq
 where the $\alpha_{(m)}$ are arbitrary constants and ${\cal L}_{(m)}^{D}$ is the $m$-th order Lanczos-Lovelock term given by,
\beq {\cal L}_{(m)}^{D} &=& \frac{1}{16\pi} \frac{1}{2^m} \delta^{a_1 b_1 \ldots a_m b_m}_{c_1 d_1 \ldots c_m d_m} R^{c_1 d_1}_{~ a_1 b_1} \cdots R^{c_m
d_m}_{~ a_m b_m} \nonumber \\
&=&  \frac{1}{16\pi} \frac{1}{2^m} \delta^{<{\bf a b (m)}>}_{<{\bf c d (m)}>} {\cal R}^{{\bf c d (m)}}_{{\bf a b (m)}}\label{LL} \eeq where $R^{c
d}_{~ a b}$ is the $D$ dimensional curvature tensor and the generalized alternating tensor $\delta^{\ldots}_{\ldots}$ is totally anti-symmetric in both set
of indices. We have used the subscript `$(m)$' to denote the order of the \LL\ theory. We have also introduced the notations, $<{\bf a b (k)}> = a_1 b_1
\ldots a_k b_k $ and \beq  {\cal R}^{{\bf c d (m)}}_{{\bf a b (m)}}= R^{c_1 d_1}_{~ a_1 b_1} \cdots R^{c_m d_m}_{~ a_m b_m}. \eeq The Einstein-Hilbert
Lagrangian is a special case of Eq.~(\ref{LL}) when $m=1$.

The Jacobson-Myers (JM) entropy associated with a stationary Killing horizon of $m$-th order \LL\ theory is \cite{Visser:1993qa, Jacobson:1993xs,
Visser:1993nu},
\beq \label{JM}
 S_H = \frac{m}{4 \hbar G} \int_s d^{D-2}y\,\alpha_{(m)} ~^{(D-2)} {\cal L}_{(m-1)}. \label{entropyofLL}
\eeq
where $^{(D-2)}{\cal L}_{(m-1)}$ is the intrinsic $(m-1)$-th order \LL\ scalar.  The JM entropy has the beautiful property that it is a function of the intrinsic horizon geometry alone.  This property is not shared by the Wald entropy (\ref{wald}), which differs from it by ambiguity terms (\ref{ambi}) involving the extrinsic curvature.

Interestingly, unlike the case of $f(R)$ gravity, the Wald entropy does not satisfy a local increase law even at the linearized order, but rather it is the JM entropy that obeys a local CSL at linearized order \cite{Kolekar:2012tq}.  However, for nonlinear processes such as black hole mergers, even the JM entropy may decrease \cite{Sarkar:2010xp}.  For the topological Lovelock terms, this decrease applies no matter how the ambiguities are fixed, and can be permanent \cite{Jacobson:1993xs, Liko, Sarkar:2010xp}.  For nontopological Lovelock terms, both the JM and Wald entropies are known to decrease instantaneously, but it is unknown whether or not the decrease can be permanent. Therefore, as long as we do not consider highly nonlinear processes like black hole mergers, the JM entropy seems to be the correct candidate for the horizon entropy of black holes in \LL\ theories.

In this article, we will demonstrate the GSL for the JM entropy, for linearized perturbations to a Killing horizon, due to minimally coupled free matter fields (or other fields satisfying the axioms of Ref. \cite{Wall:2011hj}).  Unlike the proof of the GSL for general relativity \cite{Wall:2011hj}, we will not include the effects of quantized gravitons, since this would require analyzing second order metric perturbations.\footnote{Gravitons could be consistently neglected by making a large number of matter fields, or by assuming the gravitons are in a stationary state for which the entropy does not increase.}

\subsection{f(Lovelock) gravity}

In f(Lovelock) gravity, we generalize the action still further by allowing the action to contain an arbitrary function $f$ of the set of Lovelock densities:
\begin{equation}\label{fLaction}
I = \int d^Dx \sqrt{g} \, f({\cal L}_{(1)}^{D}, \cdots {\cal L}_{(m)}^{D}),
\end{equation}
where the constant ${\cal L}_{(0)}^{D} = 1$ is omitted as redundant (a cosmological constant can already be added by choosing $f$ to have a constant piece).  f(R) and Lovelock are special cases.\footnote{Note that in $D = 4$ the Gauss-Bonnet term ${\cal L}_{(2)}^{D}$ is no longer topological when multiplied by other Lovelock densities.}

For nonlinear $f$, the equations of motion are no longer second order in the metric.  However, if $f$ is convex, there exists an  equivalent scalar formulation which is second order in the fields.  The action is given by
\begin{equation}\label{Vphi}
I = \int d^Dx\,\sqrt{g}\left[ V(\phi_1, \cdots \phi_m) + \sum \limits_{m=1}^{[D-1)/2]} \phi_m\,
{\cal L}_{(m)}^{D}\right],
\end{equation}
where $V$ is the Legendre transform of $f$.  The $\phi$ equations of motion set $\phi$ equal to functions of the Lovelock densities, allowing one to recover Eq. (\ref{fLaction}).

For f(Lovelock) gravity, we can generalize the JM entropy formula as follows:
\beq \label{JMplus}
 S_H = \frac{1}{4\hbar G} \int_s d^{D-2}y\, \! \sum \limits_{m=1}^{[D-1)/2]} \!
 m ~^{(D-2)}\!{\cal L}_{(m-1)} \frac{\partial f}{\partial {\cal L}_{(m)}}.
\eeq
This formula is no longer a function of the intrinsic horizon geometry alone.  However, it is a function of the intrinsic horizon geometry together with the Lovelock densities.

We will demonstrate that this entropy obeys a CSL and GSL for linearized perturbations to a Killing horizon.  In this case, we will need to use an additional assumption: that the background Killing horizon possesses a regular bifurcation surface.

\section{Classical Second Law}\label{csl}
Let us begin with some geometric relationships which will be useful for further discussion. In a $D$-dimensional spacetime, the event horizon is a null hyper-surface ${\cal H}$ parameterized by an affine parameter $\lambda$. The vector field $k^a = (\partial_\lambda)^a$ is tangent to the horizon and obeys affine geodesic equation. All $\lambda =$ constant
slices are space-like and foliate the horizon.  Near any point $p$ on such a slice, there is a local coordinate system $\{\lambda, x^A\}$  where $x^A, \,(A=2, \cdots ,D)$ are the coordinates of a point on $\lambda = 0$ slice connected with $p$ by a horizon generator.  We can construct a basis with the vector fields, $\{k^a, l^a,
e^{a}_{A}\}$ where $l^a$ is a second null vector such that $l^a k_a = -1$. The induced metric on any slice is $\gamma_{ab} = g_{ab} + 2 k_{(a} l_{b)}$ and
$k^a \gamma_{ab}  = 0= l^a \gamma_{ab}$. The change of the induced metric from one slice to another can be obtained from the metric evolution equation
\cite{Wald:1984rg}:
\beq
{\cal L}_{k} \gamma_{ab} = 2 \left( \sigma_{ab} + \frac{\theta}{(D-2)} \gamma_{ab} \right) \equiv 2B_{ab}, \label{metric_evolution}
\eeq
where $\sigma_{ab}$ is the shear, $\theta$ is the expansion of the horizon, and $B_{ab}$ is the null extrinsic curvature.
Then, we can obtain an evolution equation of $B_{ab}$ \cite{Gourgoulhon:2005ng}:
   \beq
{\cal L}_k B_{ab} = B_{ac} B^{c}_{b} - \gamma^{c}_{a} \gamma^{d}_{b} R_{m c n d} k^m k^n. \label{change_B}
\eeq

We would like to consider the situation when a stationary Killing horizon is perturbed by a weak fluctuation (in our work the fluctuation is semi classical) and ultimately settle down to a stationary state in the asymptotic future. Since the black hole is stationary in the asymptotic future, the vector field $k^a$ is an exact Killing vector at late times. The process is assumed to be such that all changes of the dynamical fields are first order in some suitable bookkeeping parameter $\epsilon$. More specifically, we assume that, $\theta \sim \sigma_{ab}\sim {\cal O}(\epsilon)$.
 
We shall now turn our attention to a general diffeomorphism-invariant theory of gravity described by a Lagrangian $L$. The field equation of the theory is $E_{ab} = 8\pi \, T_{ab}$ which is obtained by the metric variation of $L$. 

The entropy of the horizon should be given by a local integral of an entropy density:
\beq 
S_H = \frac{1}{4 \hbar G} \int \rho \,\,  d^{D-2}y \label{generalentropy}\eeq where $\rho$ is the entropy
density and the integration is over the $(D-2)$-dimensional space-like cross-section of the horizon. Note that, for a stationary solution, we must have $S_H = S_W$. Now, due to the accretion of matter, the entropy changes
and the change is given by
\beq\label{change_wald_ent}
\Delta S_H &=& \frac{1}{4 \hbar G}\int_{\cal H} \left(\frac{d\rho}{d\lambda} +\theta\, \rho
\right)\,d\lambda \,\, d^{D-2}y.
\eeq
We define a quantity $\Theta$ as
 \beq
  \Theta = \frac{1}{4 \hbar G} \left(\frac{d\rho}{d\lambda} +\theta\, \rho \right).
  \eeq
In case of general relativity, $\Theta$ is proportional to the expansion parameter $\theta$ of the null generators. But, in case of a general gravity theory, $\Theta$ is the rate of change
of the entropy density associated with a infinitesimal portion of horizon.  We will assume that the causal horizon eventually becomes stationary, so that $\Theta(\lambda = +\infty) = 0$.

We aim to prove that for first-order changes (in units where $G = \hbar = 1$), \beq \frac{d \Theta}{d \lambda}  \approx
- \frac{1}{4} E_{ab} k^a k^b = -2\pi\, T_{ab} k^a k^b \label{linRayGN}\eeq 
  This is our key equation.  If the causal horizons in any theory of gravity obey Eq. (\ref{linRayGN}), then this will lead to a linear CSL and semiclassical GSL. To see this explicitly, we first integrate Eq. (\ref{linRayGN}) once in the transverse directions and twice in the $\lambda$ direction \cite{Wall:2011hj}, to obtain the entropy $S_H$ at a given slice $\lambda = \lambda^\prime(y)$:
\begin{equation}\label{magic}
\frac{1}{2\pi} (S_H(\infty) - S_H({\lambda^\prime})) = \int d^{D-2}y \int_{\lambda > \lambda^\prime} \!\! T_{ab}\,k^a k^b (\lambda - \lambda^\prime) d\lambda \equiv K(\lambda^\prime).
\end{equation}
In the special case where $\lambda^\prime = 0$ is chosen to be the bifurcation surface, $K(0)$ is just the flux of Killing energy across the horizon, and therefore Eq. (\ref{magic}) is just the physical process version of the First Law: $dE = T dS$, where $E$ is the Killing energy with respect to the Killing horizon, and $T = \frac{\hbar}{2\pi}$ is the temperature.  However, by choosing $\lambda^\prime \ne 0$ the equation can be interpreted as the First Law null-translated with respect to $\lambda$.  Within the region $\lambda > \lambda^\prime$ on $H$, $K(\lambda^\prime)$ is the generator of dilations about the slice $\lambda = \lambda^\prime$ \cite{Wall:2011hj}.

Assuming the null energy condition, $K(\Lambda) \ge K(\Lambda^\prime)$ whenever $\Lambda \le \Lambda^\prime$.  The CSL follows immediately.  However, we will have to work a little bit harder to get the GSL.

\section{Generalized Second Law}\label{gsl}

Next we turn to a proof of the GSL for rapidly changing quantum matter fields.  The GSL states that $S_{H} + S_\mathrm{out}$ is an increasing function of the horizon slice, where the horizon entropy $S_{H}$ is defined as in the previous sections, and $S_\mathrm{out} = -\mathrm{tr}(\rho\,\ln\,\rho)$, $\rho$ being the state of the matter fields.  It is necessary to subtract off a state-independent UV divergence in the entanglement entropy.

In this section we will consider semiclassical fluctuations, for which the metric variation is proportional to $\hbar \sim \epsilon$.  This requires ignoring graviton fluctuations, which have $\sigma_{ab} \sim \mathcal{O}(\hbar^{1/2})$.\footnote{In general relativity, one can integrate Eq. (\ref{Ray}) at order $\hbar^{1/2}$ to show that $\theta \sim \mathcal{O}(\hbar)$, even for gravitons.  But in higher curvature gravity theories one would instead have $\Theta \sim \mathcal{O}(\hbar)$.}

We assume that the quantum fields are coupled to the gravitational fields via the semiclassical equation $E_{ab} = \langle T_{ab} \rangle$.  In the linearized approximation this equation can be derived by promoting $E_{ab} = T_{ab}$ to an operator equation and then taking the expectation values of both sides: at the linear order one can consistently identify the linearized gravitational field $\delta g_{ab}$ with its expectation value without worrying about the fact that in general $\langle AB \rangle \ne \langle A \rangle \langle B \rangle$.

Since quantum fields can violate the null energy condition, the proof requires a more detailed analysis of the matter fields.  We will assume that the matter fields have a null-surface formulation in terms of an algebra of local field operators $\mathcal{A}_H$, satisfying these axioms: (i) Local Lorentz symmetry: $\mathcal{A}_H$ contains an infinite dimensional automorphism group which includes not only the Killing symmetry but also \emph{supertranslations} $\lambda \to \lambda + a(y)$, (ii) Determinism: together with $\mathcal{I}^+$ it determines the (quantum matter data) outside of $H$, and (iii) Stability in the sense of obeying the averaged null energy condition:
\begin{equation}
\int^{+\infty}_{-\infty} T_{ab}\,k^a k^b\,d\lambda \ge 0,
\end{equation}
where the inequality is saturated by a  $|0\rangle$ which is the vacuum with respect to null-translations.  These axioms are satisfied by free fields of various spins, 1+1 CFT's, and certain kinds of superrenormalizable theories \cite{Wall:2011hj}.\footnote{Ref. \cite{Wall:2011hj} contained an additional axiom, Ultralocality, which states that the fields on different horizon generators are independent.  This axiom plays no role in the proof here, but is useful for deriving the other axioms.}

In any quantum field vacuum with a wedge region invariant under a boost symmetry, the Bisognano-Wichmann theorem ensures that the vacuum state is thermal with respect to the boost symmetry of the wedge \cite{Bisognano:1975ih}---the Unruh effect.  Sewell \cite{Sewell:1982zz} has generalized this theorem to the horizon algebra $\mathcal{A}_H$, proving that when $|0\rangle$ is restricted to the region of $H$ above $B$, it is thermal with respect to dilations about $B$ (generated by the Killing field restricted to $H$, i.e. $\xi = \lambda \hat{\lambda}$).  This implies that $|0\rangle$ is in fact the Hartle-Hawking state of the Killing manifold restricted to $\mathcal{A}_H$, whenever the Hartle-Hawking state exists (but even in cases like Kerr when it does not exist, $|0\rangle$ still exists).

By virtue of supertranslation symmetry, $|0\rangle$ is also thermal when restricted to any $\lambda > \lambda^\prime$ region:
\begin{equation}\label{thermal}
\sigma(\lambda^\prime) = \frac{e^{-2\pi K(\lambda^\prime)}}{\mathrm{tr}(e^{-2\pi K(\lambda^\prime)})},\footnote{This equation is somewhat formal since it requires renormalization to make sense.  Sewell \cite{Sewell:1982zz} proved that $\sigma$ is thermal using the more rigorous KMS definition.}
\end{equation}
where $K(\lambda^\prime)$ is the generator of dilations corresponding to the diffeomorphism $\xi(\lambda^\prime) = (\lambda - \lambda^\prime) \hat{\lambda}$, corresponding to the $T_{ab}$ integral of Eq. (\ref{magic}).\footnote{Here $K(\lambda^\prime)$ is really the \emph{canonical} dilation energy, but for minimally coupled fields this makes no difference.  In the case of a  minimally-coupled scalar field, the gravitational and canonical stress-energy tensors agree precisely.  For a spin-1/2 field, the two definitions agree when the equations of motion are satisfied.  For a spin-1 gauge field, they agree up to the equations of motion and a boundary term, but the boundary term vanishes when $\xi = 0$.  Hence it is zero at the $\lambda = \lambda^\prime$ boundary, and constant at $\lambda = +\infty$.  Hence, for purposes of proving the GSL, it is sufficient to consider the gravitational $T_{ab}$.}

This property can be used to prove that for any other semiclassical state $\rho$, the generalized entropy of $\rho$ increases as $\lambda^\prime$ is pushed forwards in time on the horizon.

The proof uses an information theoretical quantity known as the ``relative entropy'' \cite{Araki:1976zv}, which is an asymmetrical function of two states given by the following formula:
\begin{equation}
S(\rho\,|\,\sigma) = \mathrm{tr}(\rho\,\ln\,\rho) - \mathrm{tr}(\rho\,\ln\,\sigma)
\end{equation}
The relative entropy has the monotonicity property that it always decreases (or stays the same) whenever $\rho$ and $\sigma$ are both restricted to a subsystem \cite{Lindblad,Uhlmann:1976me}.  In the case of the thermal state given by Eq. (\ref{thermal}), the relative entropy can be calculated to be the difference between the free energy of $\rho$ and $\sigma$:
\begin{equation}
S(\rho\,|\,\sigma) = (2\pi K - S_{\lambda > \lambda^\prime})|_\rho - (2\pi K - S_{\lambda > \lambda^\prime})|_\sigma,
\end{equation}
where $S_{\lambda > \lambda^\prime}$ is the von Neumann entropy $-\mathrm{tr}(\rho\,\ln\,\rho)$ of the fields in the horizon algebra $\mathcal{A}_H$ restricted to the region $\lambda > \lambda^\prime$, and $K$ is understood to be an expectation value.  Using Eq. (\ref{magic}) to relate $K$ to the horizon entropy, one finds that the decrease of relative entropy as one pushes $\lambda^\prime$ forwards in time is given by
\begin{equation}
-\Delta S(\rho\,|\,\sigma) = -\Delta(2\pi K - S_{\lambda > \lambda^\prime})|_\rho = \Delta S_H(\lambda^\prime) + \Delta S_{\lambda > \lambda^\prime}) \ge 0,
\end{equation}
where $\Delta$ represents the change with time, and the free energy of $\sigma$ and $S_\infty$ have been dropped because they are the same for each slice $\sigma$.  That is, the generalized entropy of the matter on the horizon itself cannot decrease.  However, $S_{\lambda > \lambda^\prime}$ is not quite the same as the entropy outside of the black hole, because it only registers entropy that falls across the horizon.  But by the assumption of Determinism, all entropy outside of $H$ must either fall across $H$ or else escape to $\mathcal{I}^+$.  Since unitary processes preserve the entropy,
\begin{equation}
S_\mathrm{out} = S(H_{\lambda > \lambda^\prime}\,\cup\,\mathcal{I}^+) 
= S(H_{\lambda > \lambda^\prime}) + S(\mathcal{I}^+) - I(H_{\lambda > \lambda^\prime},\,\mathcal{I}^+),
\end{equation}
where $I$ is the mutual information, defined as the amount by which the entropy of two systems fails to be additive due to entanglement:
\begin{equation}\label{I}
I(A,\,B) = S(A) + S(B) - S(A\,\cup\,B).
\end{equation}
The mutual information can be defined in terms of the relative entropy as follows:
\begin{equation}
I(A,\,B) = S(\rho_{AB}\,|\,\rho_A \otimes \rho_B).
\end{equation}
Consequently it too must monotonically decrease if either $A$ or $B$ is restricted to a subregion \cite{Lindblad,Uhlmann:1976me}.  This implies that under restriction of the horizon system, the amount of entanglement with $\mathcal{I}^+$ can only decrease.  So the presence of the additional information in $\mathcal{I}^+$ can only make the generalized entropy increase faster with time \cite{sing}.  Thus the generalized second law holds:
\begin{equation}
\Delta(S_{H} + S_\mathrm{out}) \ge 0.
\end{equation}

In conclusion, we find that if a theory of gravity admits black hole solutions, such that for semiclassical states, the horizon obeys a generalized Raychaudhuri equation of the form Eq. (\ref{linRayGN}) and if certain technical conditions are satisfied by the  matter fields, the sum of the horizon entropy and the von Neumann entropy of the outside matter cannot decrease.  So, the only remaining task is to show the validity of Eq. (\ref{linRayGN}) for different theories of gravity. In the next section, we will demonstrate that black holes in both \LL\ and f(Lovelock) gravity theories indeed obey such an equation.

\section{Entropy variations in f(Lovelock)}\label{fL}
We aim to prove that for first-order changes, black holes in \fl \, gravity obey 
\beq \frac{d \Theta}{d \lambda}  \approx
 - 2\, \pi \, T_{ab} k^a k^b. 
\eeq 
As discussed in sections \ref{csl} and \ref{gsl}, the validity of this equation directly leads to the semiclassical version of the second law. As a warm up, we would first like to discuss the case of Lanczos-Lovelock gravity. 

Comparing Eq. (\ref{entropyofLL}) with Eq. (\ref{generalentropy}), we write the Jacobson-Myers (JM) entropy density associated with a stationary Killing horizon of $m$-th order \LL\ theory as
 \beq
 \rho_{(m)} =  m \, \alpha_{(m)} ~^{(D-2)}{\cal L}_{(m-1)}.\label{def_entropy_density}
\eeq
 where $^{(D-2)}{\cal L}_{(m-1)}$ is the intrinsic $(m-1)$-th order \LL\ scalar and our normalization is such that $\alpha_{(1)} = 1$.

To calculate the change of this entropy, we note that the change of the $(D-2)$-dimensional scalar $^{(D-2)}{\cal L}_{(m-1)}$ can be thought of due to the change in the intrinsic metric.
 Then, we can calculate this change by using the standard result of variation of Lanczos-Lovelock scalar. The variation of $^{(D-2)}{\cal L}_{(m-1)}$
 simply gives the equations of motion of $(m-1)$-th order Lanczos-Lovelock term in $(D-2)$ dimensions with a surface term arising due to the variation.
Therefore, for a general Lanczos-Lovelock gravity, we can write
 \beq
\frac{d\rho_{(m)}}{d\lambda} = - m \, \alpha_{(m)} \ ^{(D-2)}{\cal R}^{ab}_{(m-1)}\, {\cal L}_{k} \gamma_{ab} + D_a \delta v^a, \label{variation_section}
\eeq
\\
where $^{(D-2)}{\cal R}_{ab (m-1)}$ is the generalization of the Ricci tensor for $(m-1)$-th Lanczos-Lovelock gravity, and $D_a$ is the covariant derivative compatible with $\gamma_{ab}$. The expression for $\delta v^a$ is \cite{Mukhopadhyay:2006vu}, \beq \delta v^a = 2
\,^{(D-2)}P^{abcd}_{(m-1)}\,D_d (\delta \gamma_{bc}).
 \eeq
 For \LL\ gravity, we can ignore this surface term since the sections of the horizon are compact surfaces without boundaries, but for \fl\, it will be important.

Then using
Eq.~(\ref{metric_evolution}), we obtain,
 \begin{eqnarray}
\Theta_{(m)} =- \frac{m}{2} \, \alpha_{(m)} \biggl[\, ^{(D-2)}{\cal R}^{ab}_{(m-1)} B_{ab}  - \frac{1}{2} \theta \, ~^{(D-2)}{\cal L}_{(m-1)} \biggr].
\end{eqnarray}
Next, we refer to \cite{Kolekar:2012tq}, where it is proven that for first-order perturbations to the horizon,
\begin{eqnarray}
 \frac{ d \Theta_{(m)} }{d \lambda} = - \frac{1}{4} E_{(m)ab} k^a k^b +  {\cal O} (\epsilon^2).
\end{eqnarray}
This is entirely a geometric result and after we use the semiclassical field equation, this gives us a semiclassical version of the second law for black holes in \LL\ theory.\\

Next, we start with the extension of Jacobson-Myers entropy for \fl\, gravity given at order $m$ by
\beq S = \frac{1}{4} \int \rho \,\,  d^{D-2}y  \label{entropyGB}, \eeq where the entropy density is, \beq \rho = m \, \alpha_{(m)} \, f'({\cal L}_{(m)})
~^{(D-2)}{\cal L}_{(m-1)}  = f'({\cal L}_{(m)}) \,\rho_{(m)}.\label{def_entropy_density_fl}
 \eeq
 In this expression, in accordance with the previous section, we
have defined ${\cal L}_{(m)}$ is the $m$-th order \LL\ Lagrangian and $^{(D-2)}{\cal L}_{(m-1)}$
is the intrinsic $(m-1)$-th \LL\ scalar of the horizon cross-section.  We would like to remind the readers that as in case of \LL\ gravity, this entropy also differs from the Wald entropy $S_W$ by first-order terms.

For notational simplicity, we have written out the case when the argument of the function \fl\, is a particular $m$-th order \LL\ term, but our proof can easily be generalized for the theories whose Lagrangian is of the form $f({\cal L}_{(1)}, {\cal L}_{(2)}, \cdots)$.  One simply replaces $f'({\cal L}_{(m)})$ with $(\partial f / \partial {\cal L}_{(m)})$ and sums over $m$.

The null-null component of the field equation of \fl \, theory can be expressed as
\beq
\left[
f'({\cal L}_{(m)}) P^{abcd}\,R_{ebcd} - 2 \,P^{a}_{\,\,\,pqe} \nabla^p \nabla^q f'({\cal L}_{m})
\right] k_a k^e
= 8 \, \pi\, T^{a}_{\,\,e} k_a k^e
\eeq
where
 \beq P^{abcd}= \frac{\partial {\cal L}_{(m)}}{\partial R_{abcd} }.
 \eeq
Using the definition of ${\cal L}_{(m)}$, it is possible to show that
\beq
P^{ p q}_{r s} = \frac{m}{2^m} \delta^{p q\,< {\bf a b (m -1)}>}_{ r s\, <{\bf c d (m -1)}>}\,{\cal R}^{{\bf c d (m-1)}}_{{\bf a b (m-1)}}. \label{P_def}
 \eeq
Now, in the first order of the perturbation, we can simplify the field equation and write it in the form
 \beq
 f'({\cal L}_{m})E_{(m) ab}k^a k^b - \left( 2
\, k^a k^e \,P_{apqe} \,\gamma^{p}_{r} \gamma^{q}_{s}  + \rho_{(m)} k_r k_s \nonumber \right. \\ - \left. \, 4 \, k^a k^e \, P_{apqe} \,\gamma^{p}_{r}\,
k_s l^q \right) \, \nabla^r \nabla^s f'({\cal L}_{m})  = 8 \, \pi \, T_{ab} k^a k^b, \eeq
where $E_{(m)ab}$ is the equation of motion of $m$-th order \LL\ theory and $\rho_{(m)}$ is the corresponding entropy density. The generalized expansion parameter
of the horizon generators is given by
\beq 
\Theta = \left(\frac{d\rho}{d\lambda} +\theta\, \rho \right).
\eeq
The variation of $^{(D-2)}{\cal L}_{(m-1)}$ simply gives the equations of motion of $(m-1)$-th order Lanczos-Lovelock term in $(D-2)$ dimension and a surface terms. 
In the case of \LL\ gravity, we neglected the surface term as the cross section of event horizon is compact.  But now, 
the surface term arising from the variation of $^{(D-2)}{\cal L}_{(m-1)}$ will multiplied by $f'({\cal L}_{m})$, and is therefore no longer a total derivative.  Taking this into account, it is possible
to write the generalized expansion factor for \fl \,theory as
\beq
 \Theta =  \frac{1}{4}\left(f'({\cal L}_{m}) \Theta_{(m)} + \rho_{(m)} \frac{d}{d\lambda} f'({\cal L}_{m}) + \,4\, m\, ^{(D-2)}P^{abcd}_{(m-1)}\, B_{bc}\, D_a D_d
f'({\cal L}_{m})\right). \eeq

As in case of the \LL\ theory, here also we want to calculate the first-order change of $\Theta$. This is given by
\beq
 \frac{d \Theta}{d\lambda} \! = \! \frac{1}{4}\! \left( \! f'({\cal L}_{m}) \frac{d \Theta_{(m)}}{d\lambda} \!+\! \rho_{(m)} \frac{d^2 }{d\lambda^2} f'({\cal L}_{m}) \!+\! \,4\, m\,
^{(D-2)}P^{abcd}_{(m-1)} \, D_a D_d f'({\cal L}_{m})\frac{d B_{bc}}{d \lambda} \right). \eeq
Since $(d B_{bc} / d \lambda)$ is a first-order term, we can evaluate $D_a D_d f'({\cal L}_{m})$ on the stationary background.  This allows us to substitute
 \beq
D_a D_b f'({\cal L}_{m}) = \gamma^{c}_{a} \gamma^{d}_{b} \nabla_c \nabla_d f'({\cal L}_{m}),
 \eeq
due to the fact that $B_{ab}$ and $k^a \nabla_a f'({\cal L}_{m})$ vanish on a stationary horizon.  

At this stage, we use the result for the case of \LL\ theories, where it has been proven that
\beq
\frac{d \Theta_{(m)}}{d\lambda} = - \frac{1}{4}E_{(m) ab} k^a k^b.
\eeq
Also, we can use Eq.(\ref{change_B}) to calculate the change of $B_{bc}$.  Then, using the field equation, we finally obtain
\beq \frac{d \Theta}{d\lambda} &=& - 2\, \pi\, T_{ab} k^a k^b - \frac{1}{2}\, \left( Q^{r s} - 2
 \, k^a k^e\,P_{apqe} \,\gamma^{p r}\, k^s l^q \right) \nabla_r \nabla_s f'({\cal L}_{m}) \label{thetachange}
\eeq
where we have defined the quantity
\beq Q^{r s} = k^a k^e \left( P_{apqe} \,\gamma^{p r} \gamma^{q s} - 2\, m\, ^{(D-2)}P^{rbcs}_{(m-1)} \,R_{apqe}
\,\gamma^{p}_{b} \gamma^{q}_{c}\right). \label{Q_Def}
 \eeq


So far, we have only assumed that the background is a stationary solution. Now, we add an extra assumption that the background stationary solution also has a regular bifurcation surface.  On a stationary Killing horizon, dimensional analysis requires that (using a convention where $\lambda = 0$ at the bifurcation surface)
\beq
k^a k^e\,P_{apqe} \,\gamma^{p r}\,l^q &\propto& \frac{1}{\lambda},
\eeq
because it has one more $k$ than $l$.  So regularity requires that this expression vanish on the stationary background.  Hence the last term of Eq. (\ref{thetachange}) drops out because it is a product of two first-order terms.  (We emphasize again that Eq. (\ref{thetachange}) is only valid for first-order perturbations.)

Interestingly, for \LL\ gravity, we did not require any such regularity assumption. But, as it seems from this calculation, for \fl \, theory such an assumption is essential.  In this regard, we also note that the derivation of the first law  \cite{Iyer:1994ys,Jacobson:1993xs} is also based on the existence of a regular bifurcation surface. Moreover, if the stationary black hole solution has a regular bifurcation surface, the validity of the zeroth law (i.e. the constancy of the surface gravity on the horizon) is guaranteed independent of the equation of motion.  Hence, it is not very surprising that
 to prove the semiclassical second law for a sufficiently general theory like \fl \, gravity, we also require the same assumption.\\

We are left with 
 \beq \frac{d \Theta}{d\lambda}  = - \,2\,
\pi\, T_{ab} k^a k^b - \frac{1}{2} \, Q^{r s}\,  \nabla_r \nabla_s f'({\cal L}_{(m)}).\label{Final_form}
 \eeq

We will now show that at first order, $Q^{rs} = 0$. To prove this, let us recall the definition, $P^{ap}_{qe}$ from Eq.(\ref{P_def})
and write the first-order part of the first term in Eq.(\ref{Q_Def}) as \beq \frac{ m (m -1)}{2^{m}}k_a k^e\, \gamma_{p r} \gamma^{q}_{s}\,\delta^{a p \, <{\bf a b (m-2)}>\,g h }_{q e \,<{\bf  c d (m-2)}>\, e f}
\,{\cal R}^{{\bf c d (m-2)}\,\{0\}}_{{\bf a b (m-2)}} \, R^{ e f\,\{1\}}_{~  g h}.
 \eeq
The superscripts $\{0\}$ and
$\{1\}$ denote that the quantities are evaluated for the background and the first-order perturbation respectively. The factor $(m-1)$ is due to the fact that
$P^{ap}_{qe}$ is the product of $(m-1)$ curvatures and we have linearized the expression. Next, we also note that on the stationary background solution,
due to the antisymmetry of the alternating tensor, the curvature tensors in the above expression can have nonzero contributions only from the transverse
components and the transverse component of the full curvature tensor is equal to the intrinsic curvature of the horizon cross section. This allows us to
write \cite{Gourgoulhon:2005ng} \beq {\cal R}^{{\bf c d (m-2)}\,\{0\}}_{{\bf a b (m-2)}} =\, ^{(D-2)}{\cal R}^{{\bf c d (m-2)}\,\{0\}}_{{\bf a b (m-2)}} \eeq
Also, the alternating tensor is simply the totally anti symmetric product of $m$ Kronecker's deltas and therefore can be expressed as the determinant of a
$(m \times m)$ matrix \cite{Kothawala:2009kc}:
\begin{eqnarray}
\delta^{p q\, <{\bf a b (m -2)}> \,g h}_{ r s\, {\bf <c d (m -2)}>\, e f} = m! \, \delta^{p}_{[r} \cdots \delta^{h}_{f]}= {\mathrm{det}} \left[
\begin{array}{c|ccc} \delta^p_r & \delta^p_s & \cdots & \delta^p_{f}
\\
\hline
\\
\delta^{q}_r &  & &
\\
\vdots & & \delta^{q\,<{\bf a b (m-2)}>\, g h}_{s\,<{\bf c d (m-2)}>\, e f} &
\\
\delta^{h}_r & & &  \vphantom{\bigg{|}} \end{array} \right] . \;
\end{eqnarray}
This allows to express the $m$-th order alternating tensor in terms of the lower orders. Using all these properties, at first order in the perturbation, we
finally obtain the first-order part of the first term in Eq.(\ref{Q_Def}) as
 \beq
 && \frac{4\,
m (m -1)}{2^m} k_a k^e\, \gamma_{p r} \gamma^{q}_{s}\, \delta^{ p g\, {\bf <a b (m-2)>}  }_{q h{\bf < c d (m-2)>}}\, \,^{(D-2)}{\cal R}^{{\bf c d
(m-2)}}_{{\bf a b (m-2)}} \, R^{a g}_{e h} \nonumber \\&=& 2 m \, \gamma^{p}_{b} \gamma^{q}_{c}\,\,^{(D-2)}P^{ r b c s}_{(m-1)}\,R_{a p q e}, \eeq This
immediately shows that in the leading order $Q^{rs} = 0$ which implies
  \beq\label{result}
  \frac{d \Theta}{d\lambda}  = - 2 \, \pi\, T_{ab} k^a k^b + {\cal O}(\epsilon^2),
   \eeq
   which is what we set out to prove.
   
\section{Discussion}\label{dis}

By applying Eq. (\ref{result}) in the way suggested in sections \ref{csl} and \ref{gsl}, we have derived the linearized second law, in both its classical and generalized form, for causal horizons in f(Lovelock) gravity.  This requires us to select the correct entropy density for a general theory, namely
\beq \label{JM+}
 \rho = \sum \limits_{m=1}^{[D-1)/2]} \! m ~^{(D-2)}\!{\cal L}_{(m-1)} \frac{\partial f}{\partial {\cal L}_{(m)}}.
\eeq
This formula tells us to differentiate the action with respect to each Lovelock density, and then multiply that by the corresponding JM entropy.  It does \emph{not} agree with the Wald entropy even at the linearized level (except in the special case of f(R) gravity).

We have therefore extended the proof of the GSL, to a broad class of metric theories minimally coupled to matter fields.  As in \cite{Wall:2011hj}, the generalized entropy is shown in a differential sense, i.e. it is increasing at every point on the horizon.\footnote{Although the generalized entropy is itself a nonlocal concept, since it refers to the entire region outside of the horizon.  However, using the proof method of \cite{Wall:2011hj} it is sufficient to prove the GSL for the entropy falling across a single horizon generator.  The monotonicity of the mutual information (\ref{I}) is then sufficient to obtain the full GSL.}

However, there are some important limitations, which perhaps can be removed in future work.  First, we had to assume that the spacetime was a linearized perturbation to a Killing horizon.  This means that the entropy could differ from Eq. (\ref{JM+}) by Noether charge ambiguities (\ref{ambi}) which happen to vanish at linearized order (e.g. terms with four or more powers of $K_{ab}^i$).

It is also worth bearing in mind that linearized perturbations cannot affect the topology of a horizon.  So we also have the freedom to add or subtract a topological term to the entropy without invalidating the first law or (linearized) second laws.  This freedom was not included in the analysis of ambiguities of \cite{Iyer:1994ys, Jacobson:1993vj}, but perhaps should have been in order to avoid second law violations in topology-changing processes \cite{Sarkar:2010xp}.  In topological \LL\ gravity, the JM entropy density $~^{(D-2)} {\cal L}_{(D/2  - 1)}$ is an example of such a topological term; it is not of the form (\ref{ambi}) and can therefore affect the entropy of a stationary black hole.

Our result does not included the effect of linearized gravitons, since that would require analyzing certain \emph{second-order} variations of the metric. This is regrettable since the contribution of gravitons to the GSL is of the same order in $\hbar$ as any other matter field.  That would be a nice extension of the current result.  However, at second order the CSL/GSL can only hold for certain ranges of the coupling constants (since it is violated for general relativity with $G < 0$).

We should note that the second law has been proven with
respect to the metric horizon. In the presence of curvature, gravitons may travel along different characteristic surfaces
than light does \cite{ChoquetBruhat:1988dw, Brigante:2007nu}. For nonstationary horizons, the
location of the graviton horizon can differ from the location of the metric horizon. Since gravitons are outside the scope of our analysis, we have ignored this issue here.

We also needed to assume that the Killing horizon has a regular bifurcation surface (except in the special cases of f(R) or \LL\ gravity.)  Even in the case of black holes that form from collapse, this is a physically reasonable approximation, so long as the black hole remains stationary for a long time after the collapse, but before the entropy measurements.

In the case of the GSL, we have made some assumptions about the matter fields (see section \ref{gsl} and \cite{Wall:2011hj} for details.  It is necessary to assume that the matter fields have an algebra of observables $\mathcal{A}_H$ restricted to the horizon, satisfying certain axioms.  These axioms have been shown for free fields, superrenormalizable potential or Yang-Mills interactions, and 1+1 CFT's.  However, they appear to fail for CFT's in higher dimensions, due to the inability to form an algebra $\mathcal{A}_H$ by smearing field operators on the horizon.  A proof of the GSL in such cases may require a more delicate near-horizon limit.

Finally, although f(Lovelock) is a broad class of theories, it seems to still be an open question whether a linearized second law holds for \emph{all} metric theories of gravitation.  In such theories the action could include arbitrary covariant combinations of the Riemann tensor and its derivatives.  It is not obvious whether there is any elegant generalization of the JM-like entropy (\ref{JM+}), let alone one which is related to the equation of motion in the correct way (\ref{result})).  The problem is of importance since renormalization will typically induce quantum corrections to every possible term in the action.

Alternatively, one could look for counterexamples to the linearized second law.  Perhaps only a restricted class of theories can satisfy a semiclassical generalized second law, and other terms in the action are only allowed as perturbative corrections.

What is special about f(Lovelock) theories?  One could speculate that it is the existence of an equivalent scalar formalism (\ref{Vphi}) with second-order equations of motion.  This implies that f(Lovelock) has a null-initial data formalism on the horizon, written in terms of the metric and Lovelock densities on $H$.  One could think of this initial data as forming a classical (commuting) horizon algebra $\mathcal{A}_C$, analogous to the quantum horizon algebra $\mathcal{A}_H$ describing the matter fields.  Assuming you also know the data at $\mathcal{I}^+$ and the matter sourcing $T_{ab}$, $\mathcal{A}_C$ is sufficient to reconstruct all of the information outside of the horizon.  It is then not surprising that (\ref{JM+}) is the correct form of the entropy---it is the only candidate for the Noether-charge entropy which is contained within $\mathcal{A}_C$.  That is because it is a function of the intrinsic metric and Lovelock densities alone.

\small
\subsection*{Acknowledgements}
We are grateful for conversations with Ted Jacobson, William Donnelly, and Don Marolf.  SS's research is partially supported by the IIT Gandhinagar internal project grant: IP/IITGN/PHY/SS/2013-001.  AW is supported primarily by the Simons Foundation, with certain incidental expenses covered by grant PHY-1205500 from the National Science Foundation.

\end{document}